\documentclass[]{article}

\usepackage[affil-it]{authblk}
\usepackage[english]{babel}
\usepackage[]{natbib}
\usepackage[T1]{fontenc}
\usepackage{amsmath}
\usepackage{array}
\usepackage{txfonts}
\usepackage{graphicx,amssymb}
\usepackage[ruled,vlined]{algorithm2e}

\begin{document}
\bibliographystyle{plain}

\title{A Real-time Single Pulse Detection Algorithm for GPUs}

\author{Karel Ad\'{a}mek\thanks{Electronic address: \texttt{karel.adamek@oerc.ox.ac.uk}}\, and Wesley~Armour }
\affil{Oxford e-Research Centre, University of Oxford, Oxford, OX1 3QG, UK;}

\date{}

\maketitle

\begin{abstract}
The detection of non-repeating events in the radio spectrum has become an important area of study in radio astronomy over the last decade due to the discovery of fast radio bursts (FRBs). We have implemented a single pulse detection algorithm, for NVIDIA GPUs, which use boxcar filters of varying widths. Our code performs the calculation of standard deviation, matched filtering by using boxcar filters and thresholding based on the signal-to-noise ratio. We present our parallel implementation of our single pulse detection algorithm. Our GPU algorithm is approximately 17x faster than our current CPU OpenMP code (NVIDIA Titan XP vs Intel E5-2650v3). This code is part of the AstroAccelerate project which is a many-core accelerated time-domain signal processing code for radio astronomy. This work allows our AstroAccelerate code to perform a single pulse search on SKA-like data 4.3x faster than real-time. 
\end{abstract}

\section{Introduction}
	Single pulse detection has become significant and important in recent years due to the discovery of fast radio bursts (FRB) \citet{Lorimeretal:2005:archivalFRB}. Single pulse searches look for time isolated events (like RRATs of FRBs) but also for irregular events like giant pulses from pulsars.
	
	A single pulse detection is performed on de-dispersed time series data after de-dispersion calculated from frequency-time data recorded by a radio telescope. Single pulse detection was described by \citet{Cordes-McLaughlin:2003:ApJ:SinglePulse} as a matched filter in the form of an N-sample boxcar. In some cases (such as for zero-dm data) it is beneficial to use more targeted matched filters (for example \citet{Keane:2010:Parker} or \citet{Guidorzi:2015:MEPSA}).

\section{Single pulse detection}
We have designed and implemented a single pulse detection algorithm for NVIDIA GPUs. We use a series of $N$-sample boxcar filters, where the boxcar with width $N$ that best fits the unknown signal produces the highest SNR (signal-to-noise  ratio) which we define as $SNR = mean/standard\, deviation = \mu / \sigma$. When using boxcar filters, it is important to properly match the boxcar filter size to any signal that is present in the data otherwise we risk significant decrease in sensitivity, as was discussed in \citet{Keane:2015:FRBSensitivity}.

Our code can be considered in two parts. The first part is the event detection where we need to calculate mean $\mu$ and standard deviation $\sigma$ of the data and perform matched filtering using boxcar filters of different widths and then calculate the SNR of filtered data. This is followed by the second part: candidate selection via selecting events above some significant threshold.

\subsection{Mean and standard deviation}
There are several methods to calculate the mean $\mu$ and standard deviation $\sigma$ of a set of values. Basic algorithms to calculate the mean and standard deviation tend to be numerically unstable. A better solution was presented by \citet{Chan:1983:MSD} in form of updating (streaming) formula for the calculation of mean and standard deviation values. The serial version of this is given by equations
\begin{equation}
\label{MSD03}
\begin{aligned}
S_{1,j}& =S_{1,j-1}+\frac{1}{j\left(j-1\right)}\left(jx_j-T_{1,j}\right)\,,\\
T_{1,j}& =T_{1,j-1}+x_j\,,
\end{aligned}
\end{equation}
where $T_{1,j}$ is a value of $T$ for samples  $1 \leq n \leq j$, similarly for $S$. The mean and standard deviation are given as $\mu=T/j$ and $\sigma=\sqrt{S/j}$. In order to have a parallel algorithm we must be able to add together quantities $T$ and  $S$ which correspond to samples of different sizes. These formulae are
\begin{equation}
\label{MSD04}
\begin{aligned}
S_{1,m+n}& =S_{1,m} + S_{m+1,m+n} +\frac{m}{n\left(m+n\right)}\left(\frac{n}{m}T_{1,m} - T_{m+1,m+n}\right)^2\,,\\
T_{1,m+n}& =T_{1,m}+T_{m+1,m+n}\,.
\end{aligned}
\end{equation}
These can be calculated in parallel and have better numerical properties. The higher precision and numerical stability comes from the nature of pair-wise summation used in this algorithm.

We have chosen to separate the calculation of mean $\mu$ and standard deviation $\sigma$ into two CUDA kernels. First we divide a dataset into equal sized blocks which are then processed by the individual threadblocks of the first kernel. This produces partial results in form of $T$ and $S$ for each block using equations \eqref{MSD03} and \eqref{MSD04}. The second kernel adds these partial results together and produces estimates of the mean $\mu$ and standard deviation $\sigma$ for the dataset using equations \eqref{MSD04}.

\subsection{Approximation of standard deviation}
We can approximate the value of the standard deviation of the data after performing a boxcar by using the equation
\begin{equation}
\label{sd01}
\sigma^2_{\mathrm{A}+\mathrm{B}}=\sigma^2_{\mathrm{A}}+\sigma^2_{\mathrm{B}}\,.
\end{equation}
In our case all standard deviations are those of the same distribution giving the standard deviation after N-sample boxcar to be
\begin{equation}
\label{sd02}
\sigma_{\mathrm{N}}=\sqrt{N}\sigma\,.
\end{equation}

This approximation is not sufficient when data are dominated by a signal. In this case we use a linear approximation
\begin{equation}
\label{sd03}
\sigma_{\mathrm{N}}=\sqrt{N}\sigma\ + \left(N-1\right)\frac{\sigma_\mathrm{Nmax}}{N_\mathrm{max}-1}\,,
\end{equation}
where $\sigma_\mathrm{Nmax}$ is true value of standard deviation after applying $N_\mathrm{max}$-sampled boxcar filters. This allows us to calculate several boxcar filters with widths $1 < N \leq N_\mathrm{max}$ and an approximation of  SNR without repeating calculation for the standard deviation.

\begin{figure}[ht]
	\centering 
	\includegraphics[width=\linewidth]{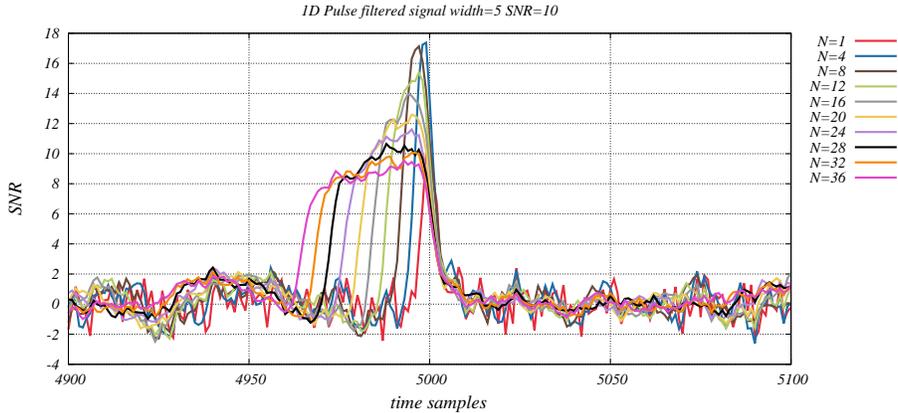}
	\caption{ The boxcar which width best fits the unknown signal produces highest signal-to-noise ratio. Here we show series of boxcar filters with different lengths and their SNR response to an input signal of fixed width.}
	\label{fig:SRAll}
\end{figure}

\subsection{Boxcar filters}
At the heart of the single pulse detection algorithm is the task of performing matched filters on each of DM trial. We perform a series of boxcar filters with different widths $1,2,\ldots,N$, where the boxcar that best fits a signal of unknown width will produce highest the SNR. This is shown in the figure \ref{fig:SRAll}. Sensitivity might be lost if boxcars are not spaced properly \citet{Keane:2015:FRBSensitivity}.

\begin{algorithm}[htp]
 \SetAlgoLined
	\SetKwData{SNR}{$\mathrm{SNR}$}
	\SetKwData{OSNR}{$\mathrm{SNR}_\mathrm{O}$}
	\SetKwData{SW}{$\mathrm{W}_\mathrm{s}$}
	\SetKwData{B}{$\mathrm{B}$}
	\SetKwData{Mean}{$\mu$}
	\SetKwData{SigmaHuh}{$\sigma$}
	\SetKwData{SigmaN}{$\sigma_\mathrm{N}$}
	\SetKwData{xel}{$x$}
	\SetKwData{lst}{$l_\mathrm{s}\!$}

	\emph{thread perform the series of boxcar filters on $X$ time samples}\;
	\For{$i=0$ \KwTo $X$}{
		\emph{setting old value of SNR to the SNR of unprocessed data}\;
		\OSNR = (\xel[\lst] - \Mean)/\SigmaHuh\;
		\SW = 1; \emph{setting detected signal width to one}\;
		\B = 0; \emph{Boxcar value}\; 
		\lst = \lst\,; \emph{starting time sample for pulse search}\;
		\For{$w=1$ \KwTo $N$}{
			\B = \B + \xel[\lst+$i$+$w$]\;
			\SNR=(\B - \Mean)/\SigmaN; \emph{\SigmaN is given by eq. \eqref{sd03}}\;
			\If{\SNR > \OSNR}{
				\OSNR = \SNR; \emph{old SRN is replaced by new one}\;
				\SW = $w$; \emph{boxcar filter width is stored}\;
			}
		}
	}
 \caption{Pseudo-code for GPU pulse detection code. The pulse detection code computes SNR for each time sample from DM trial. These are then compared with SNR of boxcar filters of different width. The highest SNR for each point is stored together with boxcar filter width.} 
\label{SPSalg}
\end{algorithm}

Implementation of boxcar filters on a GPU is straightforward. By using an approximation of the standard deviation, we do not need to recalculate the standard deviation after each boxcar filter. Thus our event detection code performs boxcar filters, calculation of SNR and finding maximum SNR for each point within one kernel. The pseudo-code of this kernel is in algorithm \ref{SPSalg}.


\subsection{Candidate selection}
	The purpose of the candidate selection is to select signals of possible astrophysical origin. The event is selected if its SNR is above some significant threshold value. This is given as a multiple of standard deviations $\sigma$. We essentially perform a following operation
		\[ x_i =
		\begin{cases}
			0       & \quad \text{if } x_i<f_\mathrm{T}\\
			x_i  & \quad \text{if } x_i \geq f_\mathrm{T}\\
		\end{cases}\,,
		\]
		where $x_i$ represents data element and $f_\mathrm{T}$ is the threshold value. 
		
	The result of the thresholding step is an unordered list of events, where neighbouring candidates are not necessary causally or spatially connected. We do not filter candidates further. The resulting list contains dispersion measure (DM) and the time at which an event was detected, its SNR and the width of the boxcar responsible for detection. 
	
\section{Results and future work}
We have developed GPU version of single pulse detection as a part of the AstroAccelerate library. Our new GPU code is 17$\times$ faster (NVIDIA Titan XP vs Intel E5-2650v3) than our OpenMP based CPU code which is already present in the library. Our current version of the AstroAccelerate code is capable of processing SKA-like data (approx. 6000 DM trials at a sampling time of $64\mu\mathrm{s}$) 4.3$\times$ faster than real-time. This includes the GPU version of the single pulse detection described here, de-dispersion by \citet{2012ASPC..461...33A} and output to a binary file. The widest boxcar filter used for measuring performance is 16 samples wide and it is performed for each point of the de-dispersed time series. In future the size of the boxcar filters will be increased, but performing each boxcar for each point is computationally expensive, hence we will use a combination of boxcar filtering with the decimation in time of the input data. 

\bibliography{P3-1}  

\end{document}